\RequirePackage{fix-cm}

\documentclass[smallextended]{svjour3}       % 
\smartqed  
\usepackage{graphicx}
\usepackage{multirow}
\usepackage{hhline}
 \journalname{General Relativity and Gravitation }
\begin{document}

\title{CRITICAL COMPACTNESS BOUND OF A CLASS OF COMPACT STARS} 



\titlerunning{CRITICAL COMPACTNESS BOUND}        % if too long for running head

\author{Satarupa Barman \and Ranjan Sharma}


\institute{Satarupa Barman\\ \email{satarupa2015@gmail.com} \\ \and  Ranjan Sharma\\ \email{rsharma@associates.iucaa.in} \at IUCAA Centre for Astronomy Research and Development, Department of Physics, Cooch Behar Panchanan Barma University, Cooch Behar 736101, West Bengal, India.}

\date{Received: date / Accepted: date}

\maketitle

\begin{abstract}
Tolman VII solution \cite{tol} is an exact analytic solution to the Einstein field equations describing the space-time of a static spherically symmetric distribution of matter. The solution has been shown to be capable of describing the interior of compact objects like neutron stars. Generalized \cite{roghunundun} and modified \cite{jiang} versions of the solution are also available in the literature, which has been subsequently developed to accommodate a wide range of neutron star EOS. The stability of the modified Tolman VII solution has recently been analyzed \cite{posada}, which provides a critical value of the adiabatic index above which the stellar configuration becomes unstable against radial oscillations. In this paper, making use of the generalized version of the Tolman VII solution, we prescribe an upper bound on the compactness ($M/R$) beyond which the star becomes unstable against radial oscillations \cite{chandra}. Our study brings to attention the role of model parameters in the generalized Tolman VII solution. The analysis also provides new insight into the role of inhomogeneity of the matter distribution vis-a-vis equation of state (EOS) on the compactness of a relativistic star.
\end{abstract}

\section{Introduction}\label{sec1}

Compact stars are unique research laboratories in the sky for gaining insight into the nature of particle interactions in extreme conditions such as extreme density, pressure and gravity. Compact stars exist either in binaries or in isolation. In the era of multi-messenger astronomy, electromagnetic and gravitational wave signals and ground-based high-energy laboratory experiments provide valuable information that can constrain the compact star equation of state (EOS). Even though systematic errors in the measurement of masses and radii of compact stars can not be ruled out \cite{miller}, observational data for the estimation of mass and radius of a compact star play crucial for gaining insight into the EOS in the ultra-high density regime of a compact star. Ideally, if the EOS is known, it is possible to numerically integrate the Tolman-Oppenheimer-Volkoff (TOV) equations to model a sequence of compact stars and subsequently, the maximum compactness for any given EOS can be obtained from the $M-R$ plot. An alternative method to study compact stars is to develop viable models for a compact star by generating exact solutions to relevant Einstein field equations. It is noteworthy that while the exterior solution to a static spherically symmetric distribution of uncharged matter is unique, namely the Schwarzschild solution, an infinite class of solutions can be obtained for the interior matter distribution. Corresponding to the Schwarzschild exterior solution, Schwarzschild obtained an interior solution \cite{arxiv}, which describes the interior geometry of a static spherically symmetric incompressible fluid distribution. Subsequently, numerous physically acceptable and well-behaved exact solutions have been developed to model compact stars like neutron and quark stars. Tolman provided one such solution \cite{tol}, which is essentially a two-parameter [$M$, $\rho_c$] family of solutions where $\rho_c$ is the central density. The analytic solution was obtained for a specific fall-off behaviour of the energy density. Recently, for better agreement with more realistic neutron star EOS, Jiang and Yagi \cite{jiang} introduced an additional parameter $\alpha$ in the density profile, thereby obtaining a three-parameter [$M$, $\rho_c$, $\alpha$] family of solutions. 

An essential physical requirement of any stellar description is its stability. Hence, it becomes imperative to investigate the impact of the model parameters on the stability of the configuration. The question we would like to address is the following - for a given stellar model, what are the bounds on the model parameters beyond which the stellar configuration becomes unstable? Chandrasekhar \cite{chandra}, in 1964, proposed a method to study the stability of a spherically symmetric stellar configuration against radial oscillations. A catalogue to solve the dynamical equation governing the stellar equilibrium under radial oscillations was later provided by Bardeen \cite{bardeen}. Many investigators have extensively used the method to examine the stability of various stellar models. For example, the technique was used by Knutsen \cite{knutsen} to analyze the stability of a sub-class of the Vaidya-Tikekar \cite{vaidya} solution describing a relativistic superdense star. The method has also been extended to the case of an anisotropic stellar body by Dev and Gleiser \cite{dev}. It has been shown that anisotropy in the core region of a stellar configuration provides more stability. Similarly, the stability of a general relativistic stellar model describing compact stars like $SAX J1808.4-3658$ was analyzed by Sharma {\em et al} \cite{sharma}. Stability analysis of the isentropic subclass of Buchdahl's exact solution was carried out Negi \cite{negi}, and the solution was found to be stable for all values of $\frac{M}{R}$ within the range $0 < \frac{M}{R} \leq 0.20$. Detailed analysis of the role of the adiabatic index on stability was performed by Moustakidis \cite{moustakidis}. Posada {\em et al} \cite{posada} have analyzed the dynamical stability of the modified Tolman VII solution \cite{tol} and obtained a  critical value of the adiabatic index at the onset of instability for specific values of the EOS parameter $\alpha$ and compactness parameter $c=M/R$. 
In a recent communication, we have reported some developments in estimating the maximum compactness bound beyond which a stellar configuration becomes unstable against radial oscillation \cite{sharma1} where different versions of the Tolman VII solution had been taken up. The current paper provides a much more detailed analysis of our investigation. Notably, the Buchdahl bound provides the maximum compactness of a relativistic star. For a homogeneous distribution of matter, the bound provides $M/R < 4/9$, as can be obtained using the Schwarzschild interior or Tolman III solutions. A more realistic description of a compact star demands a departure from homogeneity. Hence, we choose the modified Tolman VII solution for our analysis. The modified Tolman VII solution has an additional parameter which can be linked with the measure of departure from homogeneity. The parameter can also be identified as a tool to fix the EOS. In our work, making use of Chandrasekhar's method, we analyze the stability of a star for different values of the EOS parameter.  

The paper is organized as follows: In Sec.~\ref{sec2}, we lay down field equations corresponding to a static spherically symmetric relativistic compact star. The Tolman VII solution and its modified version are briefly introduced in Sec.~\ref{sec3}. In Sec.~\ref{sec4}, we outline Chandrasekhar's method to study the stability of a stellar configuration. In Sec.~\ref{sec5}, we solve the subsequent Sturm-Liouville's differential equation to determine the maximum compactness for a given EOS parameter in the Tolman VII solution. Further, making use of the causality condition in Sec.~\ref{sec6}, we make a similar study. Some concluding remarks are made in Sec.~\ref{sec7}. 

\section{Einstein field equations}
\label{sec2}

We assume the line element describing the interior of a static, spherically symmetric compact star in the standard form
\begin{equation}
ds^2 = -e^{\nu(r)}dt^2+e^{\lambda(r)}dr^2+r^2 (d{\theta^2} + \sin^2\theta d{\phi}^2). \label{1}
\end{equation}
The matter distribution inside the star is assumed to be a perfect fluid described by the energy-momentum tensor
\begin{equation}
T_{\mu\nu}=({\rho}+p)u_{\mu}u_{\nu}+p g_{\mu\nu},\label{2}
\end{equation}
where, $\rho$ is the energy density of the fluid, $p$  the isotropic pressure and $u^{\mu}$ is the $4$-velocity of the fluid.
Using equations (\ref{1}) and (\ref{2}), we obtain the Einstein field equations 
\begin{equation}
\frac{d}{dr}\left(\frac{e^{-\lambda}-1}{r^2}\right)+\frac{d}{dr}\left(\frac{e^{-\lambda}\nu'}{2r}\right)+e^{(-\lambda -\nu)}\frac{d}{dr}\left(\frac{e^{\nu}\nu'}{2r}\right) = 0 ,\label{3}
\end{equation}
\begin{equation}
 e^{-\lambda}\left(\frac{\nu'}{r}+\frac{1}{r^2}\right)-\frac{1}{r^2} = 8\pi p , \label{4} 
\end{equation}
\begin{equation}
\frac{dm(r)}{dr}=4{\pi}r^2{\rho} ,\label{5}
\end{equation}
where a prime ($'$) denotes derivative with respect to radial coordinate $r$.  The mass function $m(r)$ can be written in terms of the metric potential as
\begin{equation}
e^{-\lambda}{\equiv} 1-\frac{2m}{r}.\label{6}
\end{equation}
In the above and hereafter, we set $G=1$  and $c=1$.

\section{Original and modified Tolman VII solutions}
\label{sec3}

We note that the number of independent field equations in the previous section is less than the unknown functions, and hence it is always possible to generate an infinite class of solutions by adopting different techniques to integrate the system. However, often it is observed that not all the solutions become physically acceptable and well-behaved, as shown by Delgaty and Lake\cite{del} and Finch and Skea\cite{finch}. While the Tolman VII solution fulfils most of the necessary conditions laid down in reference \cite{del}, it has some limitations in terms of its compatibility with comparatively more realistic EOS. Hence, Raghoonundun and Hobill \cite{roghunundun} developed a generalized version of the Tolman VII solution, which can accommodate a broader range of EOS. Jiang and Yagi \cite{jiang} also developed an improved version of the Tolman VII solution to get a more realistic description of a neutron star interior. The following sub-sections outline the Tolman VII solution and some of its subsequent modifications.
  
\subsection{Original Tolman VII solution}

A stellar model is usually constructed by assuming a particular equation of state (EOS) and solving the equation of hydrostatic equilibrium, namely the Tolman-Oppenheimer-Volkoff (TOV) equations for the assumed EOS. However, Tolman \cite{tol} adopted a different technique to close the system (\ref{3})-(\ref{6}). In this technique, one assumes a particular form of one of the metric potentials given by
\begin{equation}
e^{-\lambda (r)} = 1-c{\zeta^2}(5-3{\zeta^2}),\label{7}
\end{equation}
where the parameter $c= \frac{M}{R}$ represents the stellar compactness and $\zeta=\frac{r}{R}$. The constant $R$ represents the stellar radius, and $M$ is the total mass enclosed within a radius $R$ so that $m(R)=M$. 
 
The particular choice (\ref{7}) of the metric potential $\lambda(r)$ is equivalent to choosing an energy density distribution inside the star as
\begin{equation}
\rho(r)=\rho_c (1-{\zeta}^2),\label{8}
\end{equation}
where $\rho_c$ is the central energy density of the star. 

Substitution of equation (\ref{8}) in (\ref{5}) and subsequent integration together with the regularity requirement $m(0)=0$ yield
\begin{equation}
m(r)=4\pi\rho_c \left(\frac{r^3}{3}-\frac{r^5}{5R^2}\right),\label{9}
\end{equation}
where the central density $\rho_c$ in terms of the total mass $M$ and radius $R$ is obtained in the form
\begin{equation}
\rho_c=\frac{15M}{8{\pi}R^3}. \label{10}
\end{equation}
Making use of equation (\ref{7}) in (\ref{3}) and integrating, one determines the unknown metric potential as
\begin{equation}
e^{\nu(r)} = c_1 {\cos}^2{\phi},\label {11}
\end{equation}
where
\begin{equation}
{\phi}=c_2-{\frac{1}{2}}\log\left(\zeta^2-\frac{5}{6}+\sqrt{\frac{e^{-\lambda}}{3c}}\right).\label{12}
\end{equation}
In equations (\ref{11}) and (\ref{12}), $c_1$ and $c_2$ are integration constants which can be determined from the boundary conditions (continuity of metric functions across the boundary and vanishing of pressure at the boundary)
\begin{equation}
e^{\nu(R)} = 1-\frac{2M}{R},~~~p(R)=0,~~~e^{\lambda(R)}=\frac{1}{1-\frac{2M}{R} }, \label{13}
\end{equation}
as
\begin{equation}
c_1=1-\frac{5c}{3},\label{14}
\end{equation}
\begin{equation}
c_2={\arctan}\sqrt{\frac{c}{3(1-2c)}}+\frac{1}{2}{\log}\left(\frac{1}{6}+\sqrt{\frac{1-2c}{3c}}\right).\label{15}
\end{equation}
Substitution of equation (\ref{7}) and (\ref{11}) in (\ref{4}), determines the isotropic pressure as
\begin{equation}
p = \frac{1}{4{\pi} R^2} [\sqrt{3ce^{-\lambda}}{\tan\phi}-\frac{c}{2}(5-3{\zeta}^2)]. \label{16}
\end{equation}

Thus, we have a solution expressed in terms of total mass $M$ and radius $R$. 

\subsection{Generalized Tolman VII solution}
Raghoonundun and Hobill \cite{roghunundun} extended the Tolman VII solution by considering the energy density in a more generalized form  
\begin{equation}
{\tilde{\rho}}(r)=\rho_c[1-\mu\left(\frac{r}{R}\right)^2], \label{17}
\end{equation}
where $\mu$ is a free parameter representing `stiffness' of the EOS of the star \cite{chan} whose values may vary between $0 \leq \mu\leq 1$. Note that in the extreme case of $\mu=0$, we get an incompressible fluid sphere model, and $\mu=1$ corresponds to the original Tolman VII solution. Interestingly, the $\mu=0$ case is similar to Schwarzschild's incompressible fluid sphere solution, which readily provides the maximum compactness bound $m/R < 4/9$.   

With the energy density profile (\ref{17}), equation (\ref{5}) can be integrated and applying the regularity requirement $\tilde{m(0)}=0$, one obtains
\begin{equation}
\tilde{m (r)} = 4\pi\rho_c\left(\frac{r^3}{3}-\mu\frac{r^5}{5R^2}\right),\label{18}
\end{equation}
where $\tilde{m(r)}$ is the mass function in this model. Since $\tilde{m}(R)=M$, we have
\begin{equation}
\rho_c = \frac{15M}{4{\pi}R^3(5-3\mu)}.\label{19}
\end{equation}
Using equations (\ref{3})-(\ref{6}), one obtains the unknown metric potentials as
\begin{equation}
e^{\tilde{\lambda}}=\frac{1}{1-(\frac{8\pi\rho_c}{3})r^2+(\frac{8\pi\mu\rho_c}{5R^2})r^4}=\frac{1}{1-br^2+ar^4}, \label{20}
\end{equation}
\begin{equation}
e^{\frac{\tilde{\nu}(r)}{2}}=\tilde{c_1} {\cos({\tilde{\phi}}}{\xi(r)})
+\tilde{c_2} \sin(\tilde{\phi}{\xi(r)}),\label{21}
\end{equation}
where ${\tilde{\phi}}=\sqrt{\frac{a}{4}}.$

The isotropic pressure takes the form
\begin{equation}
8{\pi}\tilde{p}=\frac{4\tilde{\phi}[\tilde{c_2}\cos(\tilde{\phi}\xi)-\tilde{c_1}\sin(\tilde{\phi}\xi)]\sqrt{1-br^2+ar^4}}{\tilde{c_1}\cos(\tilde{\phi}\xi)+\tilde{c_2}\sin(\tilde{\phi}\xi)}
-4ar^2+2b-8\pi\rho, \label{22}
\end{equation}
with
\begin{equation}
\xi(r)=\frac{2}{\sqrt{a}}\coth^{-1}\left(\frac{1+\sqrt{1-br^2+ar^4}}{r^2\sqrt{a}}\right).\label{23}
\end{equation}

The integration constants $\tilde{c_1}$ and $\tilde{c_2}$ can be determined using the boundary conditions (\ref{13}) as
\begin{equation}
\tilde{c_1}={\gamma}\cos(\tilde{\phi}\xi_R)-\frac{\alpha}{\tilde{\phi}}\sin(\tilde{\phi}\xi_R),\label{24}
\end{equation}
\begin{equation}
\tilde{c_2}={\gamma}\sin(\tilde{\phi}\xi_R)+\frac{\alpha}{\tilde{\phi}}\cos(\tilde{\phi}\xi_R).\label{25}
\end{equation}

The advantage of this solution is that the stiffness parameter $\mu$ can be suitably adjusted according to the requirement of a given EOS. Consequently, in this paper, we intend to take up this solution to analyze how the stiffness factor $\mu$ vis-a-vis EOS influences the stability of a stellar composition. 

\subsection{Improved Tolman VII solution}
Jiang and Yagi \cite{jiang} have separately proposed an improved version of the Tolman VII solution for a more realistic description of neutron star interiors. In their approach, the energy density is assumed to be of the form 
\begin{equation}
\rho(r)=\rho_c[1-\alpha{\zeta}^2+(\alpha-1){\zeta}^4],\label{eq26}
\end{equation}
where $\rho_c$ is the central density, $R$ is the stellar radius, $\alpha$ is a free parameter which should be fixed so that the condition $\rho(R)=0$ is always satisfied. Note that $\alpha\rightarrow 1$ represents the original Tolman VII solution. 

Posada {\em et al} \cite{posada} examined the dynamical stability of stellar configurations using this particular solution. 

\section{Stability: Chandrasekhar's method}
\label{sec4}
Chandrasekhar, in 1964, introduced the variational method to analyze the stability of a spherically symmetric star against radial oscillations. To obtain an upper bound on the physical variables beyond which instability might develop, in our calculation, we follow the same technique as proposed by Chandrasekhar. In this approach, for a spherically symmetric radially oscillating star, one assumes the line element in the standard form  
\begin{equation}
ds^2 = -e^{\nu(r,t)}dt^2+e^{\lambda(r,t)}dr^2+r^2 (d{\theta^2} + \sin^2\theta d{\phi}^2),\label{26}
\end{equation}
where
\begin{eqnarray}
\nu(r,t)&=&\nu_0(r)+\delta\nu(r,t),\label{27}\\ 
\lambda(r,t)&=&\lambda_0(r)+\delta\lambda(r,t),\label{28}
\end{eqnarray}
are functions of $r$ and $t$. In (\ref{27}) and (\ref{28}), $\nu_0(r)$ and $\lambda_0(r)$ represent metric potentials corresponding to the equilibrium configuration and $\delta\nu(r,t)$ and $\delta\lambda(r,t)$ represent functions due to perturbations from the equilibrium configuration. The physical variables like energy density and pressure are accordingly written as
\begin{eqnarray}
\rho(r,t)&=&\rho_0(r)+\delta\rho(r,t),\label{29}\\
p(r,t)&=&p_0(r)+\delta p(r,t).\label{30}
\end{eqnarray}
The radial perturbation of the fluid from its equilibrium position is assumed to be of the form
\begin{equation}
\delta r=\frac{u_n(r)}{r^2} e^{\frac{\nu_0(r)}{2}} e^{i\omega_n t},\label{31}
\end{equation}
where $u_n(r)$ is the amplitude of $n$th mode of radial oscillation and $\omega_n$ is the characteristic frequency of oscillation. 

The dynamical equation governing the stellar oscillation can be expressed in standard Sturm-Liouville's differential equation form
\begin{equation}
P(r)\frac{d^2u_n(r)}{dr^2}+\frac{dP}{dr}+[Q(r)+{\omega_n}^2W(r)]u_n(r)=0, \label{32}
\end{equation}
where,
\begin{equation}
P(r)=\gamma p_0e^\frac{(\lambda_0+3\nu_0)}{2}r^{-2},\label{33}
\end{equation}
\begin{equation}
Q(r)=e^\frac{(\lambda_0+3\nu_0)}{2}\left[\frac{{p_0'}^2}{r^2(p_0+\rho_0)}-\frac{4p_0'}{r^3}-\frac{{8\pi}p_0}{r^2}(\rho_0+p_0)e^{2\lambda_0}\right],\label{34}
\end{equation}
and
\begin{equation}
W(r)=e^\frac{(3\lambda_0+\nu_0)}{2}r^{-2}(\rho_0+p_0).\label{35}
\end{equation}

For the fundamental mode of oscillations, the pulsation equation takes the form 
\begin{eqnarray}
&&\omega_0^2 \int^{R}_{0}exp\left[\frac{1}{2}(3\lambda_0+\nu_0)\right](p_0+\rho_0)\frac{u^{2}_{0}}{r^2}dr = \nonumber \\
 &&\int^{R}_{0}exp\left(\frac{1}{2}(3\nu_0+\lambda_0)\right)\left(\frac{p_0+\rho_0}{r^2}\right) 
 \left([-\frac{2}{r}+\frac{d\nu_0}{dr}-\frac{1}{4}(\frac{d\nu_0}{dr})^2+8{\pi}p_0 exp(\lambda_0)]u^2\right. \nonumber \\
 &&\left.+\frac{dp_0}{d\rho_0}\left(\frac{du_0}{dr}\right)^2\right)dr,  \label{36}
\end{eqnarray}
where the adiabatic index is defined as $\gamma=\frac{p+\rho}{p}\frac{dp}{d\rho}.$

A relativistic stellar model will be stable against radial oscillations if the fundamental frequency of oscillation is real and positive. In the following section, we perform numerical integration of equation (\ref{36}) for specific stellar configurations to evaluate values of the fundamental frequencies. 

To solve (\ref{36}), we employ the method 2D given by Bardeen et al\cite{bardeen} and assume a trial solution with the following boundary conditions:
\begin{itemize}
  \item (i) Fluid at the star's centre is not displaced during the radial oscillation.
      
  \item (ii) The Lagrangian change in pressure ($\Delta{p}$) at the surface ($r=R$) must vanish which implies
         \begin{equation}
             \frac{du_0}{dr}\rightarrow{0}~~~ as~~~ r\rightarrow{R} .\label{38a}
          \end{equation}
\end{itemize}

In our calculation, we choose the trial function in the form
\begin{equation}
u_0=b_1 r^3 + b_2 r^5, \label{38}
\end{equation}
which satisfies the boundary condition (i). Boundary condition (ii) restricts the choice of the constants $b_1$ and $b_2$ by the relation 
\begin{equation}
3 b_1 R^2 + 5 b_2 R^4 = 0.\label{39}
\end{equation}
Note that the constants $b_1$ and $b_2$ can take values ranging from $-\infty$ to $+\infty$. For example, in Table~\ref{tab1}, we have tabulated the values of $b_2$ for different choices of $b_1$ where we have considered a star of radius $R=10~$km.
\begin{table}[ht]
\caption{Values of $b_2$ for different choices of $b_1$ satisfying condition (\ref{39}) where we have assumed a star of radius $R=10~$km.}\label{tab1}
\begin{tabular}{|c|c|c|}
\hline
$b_1$~(in km$^{-2}$) & $b_2$~ (in km$^{-4}$)  \\
\hline
$1$    & $-6\times10^{-3}$ \\
100 & -0.6\\
$10^{10}$    & $-6\times 10^7$\\
-100 & 0.6\\
$-10^{10}$    & $6\times 10^7$\\
\hline
\end{tabular}
\end{table}

\section{Critical compactness in generalized Tolman VII solution}
\label{sec5}
\begin{figure}[ht]%
\centering
\includegraphics[width=0.9\textwidth]{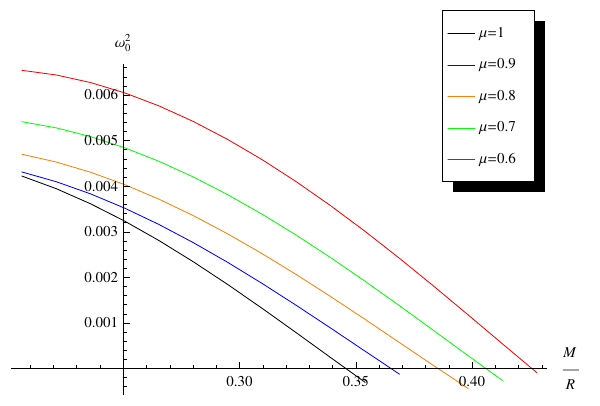}
\caption{Variation of the fundamental frequency ($\omega_0$) with maximum compactness ($\frac{M}{R}$) for different values of EOS parameter $\mu$. The compactness has been varied by varying the mass for a fixed radius or by varying the radius keeping the mass fixed for different values of $\mu$. The corresponding zero value of the fundamental frequency determines the maximum compactness bound. Beyond the maximum compactness bound, the fundamental frequency becomes negative, implying the development of instability. We note that the critical compactness decreases with increasing values of stiffness parameter $\mu$.}\label{fig1}
\end{figure}

To evaluate the maximum compactness bound, using a generalized Tolman VII solution, we numerically integrate equation (\ref{36}) for specific stellar configurations and evaluate the values of the fundamental frequencies ($\omega_0^2$). For a given radius and fixed value of the EOS parameter $\mu$, the total mass is increased till the \textbf{fundamental frequency} becomes a \textbf{negative quantity}. Similarly, the mass is kept fixed, and the radius is decreased till the configuration becomes unstable. The procedure, thus, determines the maximum compactness bound $(M/R)$ for a given value of the stiffness parameter $\mu$. In our analysis, we have not considered the limiting case $\mu=0$, which represents a constant-density star. However, for other values of $0< \mu \leq 1$, $\omega_0^2$ has been evaluated, and the results are compiled in Table~\ref{tab2}. The Table shows the fundamental frequency at the onset of instability and related parameters. The bold values indicate the maximum compactness bounds. The variation of $\omega_0^2$ with compactness for a given value of $\mu$ is shown in Fig.~\ref{fig1}. The plot is an outcome of the results obtained in Table~2. It shows that the critical compactness decreases with increasing values of $\mu$, which denotes a departure from a constant-density star. We have considered a wide range of values of the constants $b_1$ and $b_2$ (both negative and positive). It turns out that the critical value of the fundamental frequency $\omega_0$ remains independent of the choice of the constants appearing in the assumed trial wave function, which implies that the corresponding critical compactness does not depend on the choice of the constants of the assumed trial function. This is shown in Table~\ref{tab3}.

\begin{table}[h]
\caption{Maximum compactness bound and corresponding central density beyond which the stellar configuration becomes dynamically unstable in the case of generalized Tolman VII solution. By keeping the radius fixed at $R=10~$km and increasing the total mass, we have obtained the maximum compactness beyond which the fundamental frequency becomes a negative quantity. The bold value indicates the maximum compactness bound for a given stiffness parameter $\mu$.}
\label{tab2}
\begin{tabular}{|c|c|c|c|c|}
\hline
$\mu$ & Mass (M) ($M_\odot$) & Compactness(M/R) & Central density (gm~cm$^{-3}$) & Fundamental frequency ($\omega_0^2$)  \\
\hline
 \multirow{4}{*}{0.4}  & 1.4			& 0.2065			& $8.7525\times10^{14}$ & 0.0109\\
      & 2.9		& 0.42775			& $1.81303\times10^{15}$ & 0.0022\\
    & \textbf{3}			& \textbf{ 0.4425}	&		$\textbf{1.8755}\times10^{15}$ & \textbf{0.0003} \\
     & 3.1 & 0.45725 & $1.93806\times10^{15}$ & -0.0027 \\
    \hline
		\multirow{4}{*}{0.5} & 1.4			& 0.2065			& $9.5027\times10^{14}$ & 0.0083\\
        & 2.5		& 0.36875			& $1.6969\times10^{15}$ & $0.0038$ \\
        & \textbf{2.9}			& \textbf{0.42775}			& $\textbf{1.9890}\times10^{15}$ &$\textbf{0.0008}$\\ 
       & 3 & 0.4425 & $2.03631\times10^{15}$ & -0.0010\\
  \hline 
\multirow{4}{*}{0.6}	& 1.4			& 0.2065			& $1.0393\times10^{15}$ & 0.0065\\
     & 2.5 		& 0.36875 & $1.8560\times10^{15}$ & 0.0024\\
    & \textbf{2.8}			& \textbf{0.413}			& $\textbf{2.0787}\times10^{15}$ &\textbf{0.0005}\\
    & 2.9 & 0.42775 & $2.1527\times10^{15}$ & -0.0005\\
     \hline 
     \multirow{4}{*}{0.7} & 1.4		& 0.265 & $1.1468\times10^{15}$ & 0.0054\\
     & 2.5		&  0.36875			& $2.0480\times10^{15}$ & 0.0014\\
       & \textbf{2.7}			& \textbf{0.39825}		& $\textbf{2.2118}\times10^{15}$ &\textbf{ 0.0003}\\
      & 2.8 & 0.413 & $2.29377\times10^{15}$ & $-0.0002$\\
      \hline 
		\multirow{4}{*}{0.8} 	 & 1.4			& 0.2065			& $1.2792\times10^{15}$ & 0.0046\\
       & 2.5			& 0.36875			& $2.2843\times10^{15}$ & 0.0006\\
       & \textbf{2.6}			& \textbf{0.3835}			& $\textbf{2.37569}\times10^{15}$ & \textbf{0.00006}\\
       & 2.7 & 0.3925 & $2.4670\times10^{15}$ & -0.0004\\
        \hline
	\multirow{3}{*}{0.9}	 & 1.4			& 0.2065		& $1.4460\times10^{15}$ & 0.0043 \\
        & \textbf{2.4} & \textbf{0.354} & $\textbf{2.4789}\times10^{15}$ & \textbf{0.0004}\\
       & 2.5			& 0.36875			& $2.5822\times10^{15}$ & -0.00012\\
    \hline 
	\multirow{3}{*}{1}	 & 1.4			& 0.2065			& $1.6639\times10^{15}$ & 0.0042\\ 
         & \textbf{2.3}			& \textbf{0.33925}			& $\textbf{2.7320}\times10^{15}$ & \textbf{0.0002}\\ 
        & 2.4			& 0.354			& $2.8508\times10^{15}$ & -0.0003 \\
         \hline
	
				\end{tabular}
\end{table}

\begin{table}
\caption{variation of the fundamental frequency for different choices of the constants $b_1$ and $b_2$ in the assumed trial wave function. We assumed a star of fixed radius $R=10~$km. }
\label{tab3}
	\centering
		\begin{tabular}{|c|c|c|c|c|c|}
		\hline
		$\mu$ & Mass($M_\odot$) & Compactness($\frac{M}{R}$) & $b_1$~(km$^{-2}$) & $b_2$~ (km$^{-4}$) & Fundamental frequency ($\omega_0^2$)  \\
		\hline
		\multirow{5}{*}{0.4} & \multirow{5}{*}{3} & \multirow{5}{*}{0.4425} & 1 & $-6\times10^{-3}$ & 0.0003 \\
		\hhline{~~~---}
	   & & &  $10^2$ & $-0.6$  & 0.0003 \\
	   \hhline{~~~---}
	   & & & $10^{10}$ & $-6\times10^7$ & 0.0003 \\
	   \hhline{~~~---}
	   & & & $-10^2$ & $0.6$ & 0.0003 \\
	   \hhline{~~~---}
	   & & & $-10^{10}$ & $6\times10^7$ & 0.0003\\
	   \hline
	   \multirow{5}{*}{0.6} & \multirow{5}{*}{2.8} & \multirow{5}{*}{0.413} & 1 & $-6\times10^{-3}$ & 0.0005\\
	   \hhline{~~~---}
	   & & & $10^2$ & $-0.6$  & 0.0005 \\
	   \hhline{~~~---}
	   & & & $10^{10}$ & $-6\times10^7$ & 0.0005 \\
	   \hhline{~~~---}
	   & & & $-10^2$ & $0.6$ & 0.0005 \\
	   \hhline{~~~---}
	   & & & $-10^{10}$ & $6\times10^7$ & 0.0005 \\
	   \hline
	   \multirow{5}{*}{1} & \multirow{5}{*}{2.3} & \multirow{5}{*}{0.33295} & 1 & $-6\times10^{-3}$ & 0.0002\\
	   \hhline{~~~---}
	   & & & $10^2$ & $-0.6$  & 0.0002 \\
	   \hhline{~~~---}
	   & & & $10^{10}$ & $-6\times10^7$ & 0.0002 \\
	   \hhline{~~~---}
	   & & & $-10^2$ & $0.6$ & 0.0002 \\
	   \hhline{~~~---}
	   & & & $-10^{10}$ & $6\times10^7$ & 0.0002 \\
	   \hline			
		\end{tabular}
\end{table}

In Table~\ref{tab4}, we have compiled the values of the maximum compactness bound for different $0< \mu \leq 1$. Note that in all the cases, we get $\frac{2M}{R} < 1$. However, for values $\mu < 0.4$, the procedure yields critical compactness, which is more than the Buchdahl compactness bound $(\frac{M}{R})\leq \frac{4}{9}$. The critical compactness for $\mu=1$ is found to be $0.33925$, which is exactly the same as obtained earlier by Posada {\em et al} \cite{posada}. Interestingly, the maximum compactness bound decreases with the increase of inhomogeneity vis-a-vis stiffness parameter $\mu$. This makes sense as the most compact object is a constant-density star that cannot be compressed further. The variation of critical compactness with $\mu$ is shown in Fig.~2. 

\begin{figure}[ht]%
\centering
\includegraphics[width=0.9\textwidth]{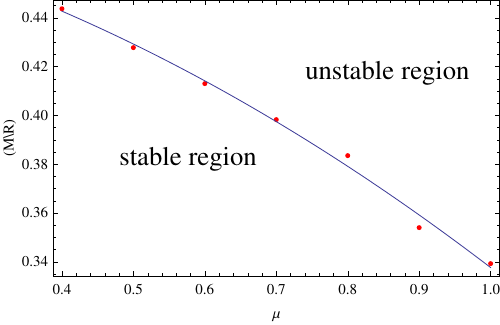}
\caption{Upper bound on compactness for different values of $\mu$ in the generalized Tolman VII solution. Here the red dots represent the critical values of the compactness for stability, which we get after numerically integrating equation (\ref{36}) for a particular $\mu$. A parabolic curve represented by $M/R= 0.48043-0.06202\mu - 0.08077 {\mu}^2$ fits well with our data, which is represented by the blue curve. For all the values of the compactness below the curve, the model is stable against radial oscillations.}\label{fig2}
\end{figure}

To understand the relationship between the maxium compactness bound ($\frac{M}{R}$) and the inhomogeneous prameter ($\mu$), utilizing Fig.~\ref{fig2}, we obtain the corresponding best-fit curve 
$$\frac{M}{R}= a_1 + a_2\mu + a_3\mu^2,$$
where $a_1=0.48043$, $a_2= -0.06202$ and $a_3=- 0.080774$. 

\begin{figure}[ht]%
\centering
\includegraphics[width=0.9\textwidth]{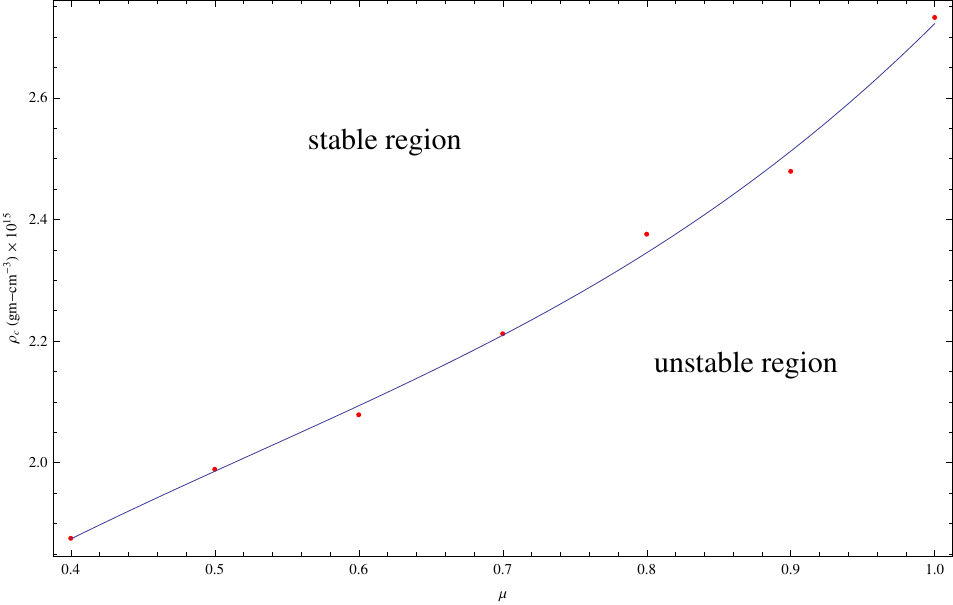}
\caption{Permissible maximum central density for stability for different values of $\mu$ in the generalized Tolman VII solution. Here the red dots represent the critical values of the central density for stability. A parabolic curve represented by $\rho_c =1.16192\times10^{15} + 2.70222\times10^{15} \mu - 3.07624\times10^{15} {\mu}^2+ 1.93309\times10^{15} {\mu}^3)$ fit well with our data, which is represented by the blue curve. For all the values of the central density below the curve, the model is stable against radial oscillations.}\label{fig3}
\end{figure}

The maximum central density corresponding to the maximum compactness in terms of $\mu$ is obtained as 
$$\rho_c = \frac{15M}{4{\pi}R^3(5-3\mu)}.$$
In Fig.~\ref{fig3}, we show graphically the maximum permissible central density of a star of a particular radius (in our calculation, we assume $R=10~$km) for different values of the EOS parameter $\mu$. The plot separates the stable stellar configurations from unstable configurations in terms of central density and EOS parameter $\mu$. We note that higher values of $\mu$ permit a higher value of the central density. 
The best-fit curve, in this case, is obtained in the form
$$\rho_c= c_1+c_2\mu + c_3\mu^2 +c_4\mu^3,$$
where $c_1=1.16192\times10^{15}$, $c_2=2.70222\times10^{15}$, $c_3=- 3.07624\times10^{15}$ and $c_4=1.933\times10^{15}$ in units gm~cm$^{-3}$.

\begin{table}[ht]
\caption{Variation of the critical compactness $(M/R)_{max}$ and maximum central density$(\rho_c)$ with inhomogeneous parameter $(\mu)$.}\label{tab4}%
\begin{tabular}{|c|c|c|}
\hline
$\mu$ & $(M/R)_{critical}$  & central density $(\rho_c)$  \\
& & (gm~cm$^{-3}$) \\
\hline
%\midrule
$0.4$ & $0.4425$ & $1.8755\times10^{15}$ \\
$0.5$ & $0.4277$ & $1.9890\times10^{15}$ \\
$0.6$ & $0.4130$ &  $2.0787\times10^{15}$\\
$0.7$ & $0.3982$ & $2.2118\times10^{15}$ \\
$0.8$ & $0.3835$ & $2.3756\times10^{15}$ \\
$0.9$ & $0.3540$ & $2.4789\times10^{15}$\\
$1$ & $0.3392$ & $2.7320\times10^{15}$\\
\hline
%\botrule
\end{tabular}
\end{table}

\section{Speed of sound within a fluid and bounds on compactness based on causality condition}
\label{sec6}
\begin{figure}[ht]%
\centering
\includegraphics[width=0.8\textwidth]{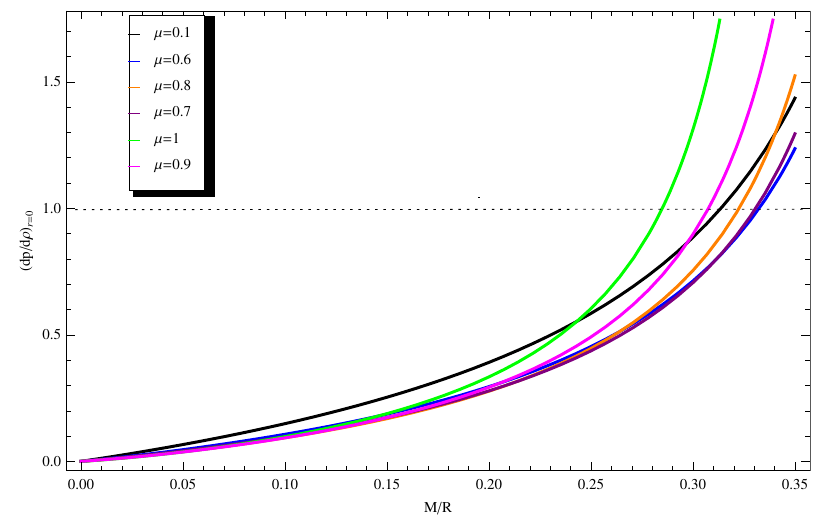}
\caption{Speed of sound at the centre as a function of compactness $(\frac{M}{R})$ for different values of the parameter $\mu$. The dotted line indicates the upper bound on compactness restricted by the causality condition $\frac{ dp}{ d\rho} \leq 1$.}\label{fig4}
\end{figure}

The speed of sound inside a star is an important property that needs to be probed to develop a stellar model. The speed of sound is given by the relation $v_s^2=(\frac {dp}{d\rho})$. The causality condition demands that the speed of sound $(\frac {dp}{d\rho}) \leq 1$. Now, the speed of sound is maximum at the centre of a star, and hence we find it worthwhile to examine the causality condition at the centre of the star. In Fig.~\ref{fig4}, we show the variation of the speed of sound $(\frac{dp}{d\rho})$ at the centre as a function of compactness $(\frac {M}{R})$ for different values of the stiffness parameter $\mu$. Obviously, the dotted line in the plot provides the upper bound on the compactness beyond which the causality condition is violated. We note that the upper bound on compactness decreases with increasing values of $\mu$. This is shown in Table~\ref{tab5}, which suggests that as one moves away from homogeneous distribution, the compactness decreases. 

\begin{table}[h]
\caption{Maximum compactness obtained by employing the causality requirement for different values of the EOS parameter $\mu$.}\label{tab5}%
\begin{tabular}{|c|c|c|}
\hline
$\mu$ & $(\frac{M}{R})_{max}$  \\
\hline
$0.6$    & $0.3235$     \\
$0.7$    & $0.3195$     \\
$0.8$    & $0.3104$    \\
$0.9$    & $0.2942$     \\
$1$      & $0.2709$    \\
\hline
\end{tabular}
\end{table}

\begin{figure}[ht]%
\centering
\includegraphics[width=0.8\textwidth]{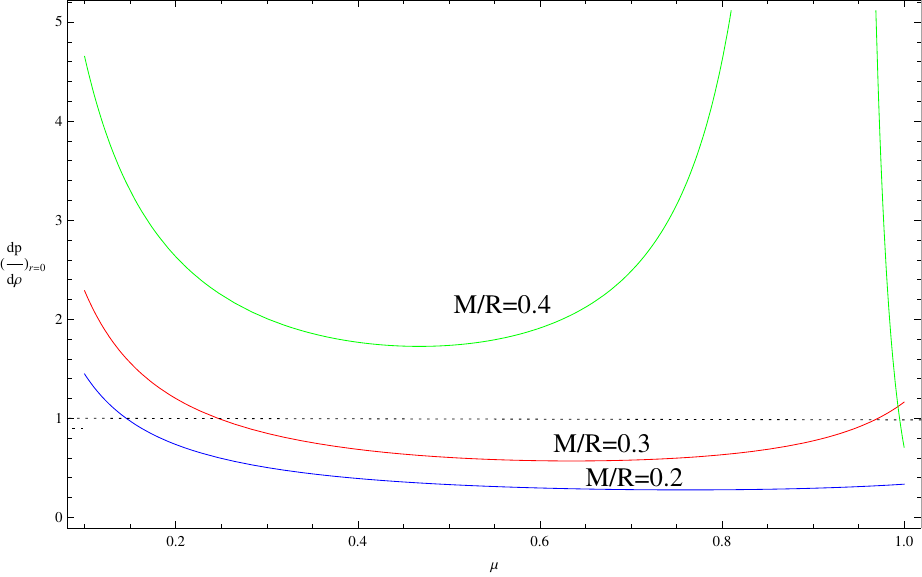}
\caption{Speed of sound at the centre as a function of parameter $(\mu) $ for different compactness values. The dotted line indicates the causality condition $\frac{ dp}{ d\rho} \leq 1$.}\label{fig5}
\end{figure}

In Fig.~\ref{fig5}, we plot the central values of the sound speed $v_s^2=(\frac{dp}{d\rho})$ as a function of EOS parameter $\mu$ for different compactness. Interestingly, the causality condition seems to get violated as one chooses higher compactness values. Moreover, the plot provides a bound on the stiffness parameter $\mu$ governed by the causality condition for a given compactness. This also rules out the unrealistic case of a constant density star with $\mu=0$ where the sound speed exceeds the upper limit even for comparatively smaller values of compactness.

Finally, in Table (\ref{tab6}), we show a comparative study of the critical compactness based on causality condition ($\frac{dp}{d\rho} \leq 1$) and stability against radial oscillations. The Table shows that the causality condition puts a more stringent bound on compactness than the results based on stability analysis.

\begin{table}[ht]
\caption{Comparison between the limiting values of compactness based on causality requirement and stability against radial oscillations.}\label{tab6}%
\begin{tabular}{ |c|c|c|  }
 \hline
  & \multicolumn{2}{|c|}{$(M/R)_{critical}$} \\
 \hline
 $\mu$ & (Based on stability analysis) & $v_s^2(0)\leq1$ \\
  \hline
$0.6$ & $0.413$ &  $0.3389$\\
$0.7$ & $0.39825$ & $0.3330$ \\
$0.8$ & $0.3835$ & $0.3241$ \\
$0.9$ & $0.354$ & $0.3093$\\
$1$ & $0.33925$ & $0.279$\\
\hline
\end{tabular}
\end{table}
\section{Concluding Remarks}

\label{sec7}

Our analysis shows that a departure from homogeneous matter distribution, characterized by the model parameter $\mu$ in the Tolman VII solution, is crucial in fixing the maximum compactness bound beyond which the star becomes dynamically unstable. The most compact object has a homogeneous distribution of matter. As inhomogeneity in the matter distribution sets in, the upper bound on compactness decreases. The subsequent maximum central density, in contrast, increases. It is to be stressed that the critical bound for the generalized Tolman VII solutions never exceeds the black hole limit $M/R < 1/2$. In our calculation, the maximum compactness bound was obtained in two different ways: (i) For a given radius, the total mass was increased till the configuration became unstable, and (ii) for a fixed mass star, the boundary was decreased till the configuration became unstable. In both cases, we obtain the same compactness bound, as expected. We trust that this procedure has never been adopted in earlier analyses. We also obtain the maximum compactness bound by employing the requirement that the sound speed (at the centre) should be causal. Fulfillment of the causality condition at the centre provides a more stringent compactness bound as compared to its value obtained by analyzing the stars' stability against radial oscillations.  

In the Tolman VII solution, since the parameters $\mu$ can also be linked with EOS, our results clearly show an intricate relationship between the maximum compactness bound and the EOS. We have generated a polynomial relationship between the EOS parameter $\mu$ and the maximum compactness bound, which is likely to provide new insight into the physics of compact objects.

\section*{Acknowledgments}
We express our sincerest thanks to the anonymous referee for providing valuable suggestions.

RS gratefully acknowledges support from the Inter-University Centre for Astronomy and Astrophysics (IUCAA), Pune, India, under its Visiting Research Associateship Programme.

\section{Declarations}

\begin{itemize}
\item Funding:~Not Applicable
\item Conflict of interest/Competing interests:~The authors declare no conflict of interest.
\item Ethics approval 
\item Consent to participate:~All authors have read and agreed to the published version of the manuscript.
\item Consent for publication:~All authors have read and agreed to the published version of the manuscript.
\item Availability of data and materials:~The data underlying this article is available in the public domain as cited in the references.
\item Code availability 
\item Authors' contributions:~Conceptualization, methodology, numerical computation, validation, formal analysis, investigation, resources,  data curation, writing---original draft preparation: RS and SB; \\Supervision: RS
\end{itemize}

\end{document}